\documentclass[prd,superscriptaddress,nofootinbib,amsmath,amssymb,aps,10pt]{revtex4}

\usepackage{bm}
\usepackage{amsfonts}
\usepackage{latexsym}
\usepackage[latin1]{inputenc}
\usepackage{graphicx}
\usepackage{amsmath}
\usepackage{palatino}
\usepackage{mathpazo}
\usepackage[outdir=./]{epstopdf}

\usepackage{booktabs}
\usepackage{dcolumn}

\def\jnl@style{\it}
\def\aaref@jnl#1{{\jnl@style#1}}

\def\aaref@jnl#1{{\jnl@style#1}}

\def\aj{\aaref@jnl{AJ}}                   
\def\apj{\aaref@jnl{ApJ}}                 
\def\apjl{\aaref@jnl{ApJ}}                
\def\apjs{\aaref@jnl{ApJS}}               
\def\apss{\aaref@jnl{Ap\&SS}}             
\def\aap{\aaref@jnl{A\&A}}                
\def\aapr{\aaref@jnl{A\&A~Rev.}}          
\def\aaps{\aaref@jnl{A\&AS}}              
\def\mnras{\aaref@jnl{Mon.~Not.~Roy.~Astron.~Soc.}}             
\def\prd{\aaref@jnl{Phys.~Rev.~D}}        
\def\prc{\aaref@jnl{Phys.~Rev.~C}}  
\def\prl{\aaref@jnl{Phys.~Rev.~Lett.}}    
\def\qjras{\aaref@jnl{QJRAS}}             
\def\skytel{\aaref@jnl{S\&T}}             
\def\ssr{\aaref@jnl{Space~Sci.~Rev.}}     
\def\zap{\aaref@jnl{ZAp}}                 
\def\nat{\aaref@jnl{Nature}}              
\def\aplett{\aaref@jnl{Astrophys.~Lett.}} 
\def\apspr{\aaref@jnl{Astrophys.~Space~Phys.~Res.}} 
\def\physrep{\aaref@jnl{Phys.~Rep.}}      
\def\physscr{\aaref@jnl{Phys.~Scr}}       
\def\commat{\aaref@jnl{Comm.~Math.~Phys.}}              
\def\science{\aaref@jnl{Science}}               
\def\cqg{\aaref@jnl{Classical Quant.~Grav.}}            
\def\jpcs{\aaref@jnl{JPCS}}                                     
\def\ijmpd{\aaref@jnl{Int.~J.~Mod.~Phys.~D}}                    
\def\grg{\aaref@jnl{Gen.~Relat.~Gravit.}}               
\def\rpp{\aaref@jnl{Rep.~Prog.~Phys.}}          
\def\npa{\aaref@jnl{Nucl.~Phys.~A}}        
\def\lrr{\aaref@jnl{Living Rev.~Rel.}}                   
\def\jcap{\aaref@jnl{J.~Cosmology Astropart.~Phys.}}    
\def\rmp{\aaref@jnl{Rev.~Mod.~Phys.}}   
\def\rmp{\araa@jnl{Annu.~Rev.~Astron.~Astrophys.}}   

\allowdisplaybreaks[1]

\begin{document}

\title{Tidal Love numbers of neutron stars in $f(R)$ gravity}

\author{Stoytcho S. Yazadjiev}
\email{yazad@phys.uni-sofia.bg}
\affiliation{Department
	of Theoretical Physics, Faculty of Physics, Sofia University, Sofia
	1164, Bulgaria}
\affiliation{Theoretical Astrophysics, Eberhard-Karls University
	of T\"ubingen, T\"ubingen 72076, Germany}
\affiliation{Institute of Mathematics and Informatics, 	Bulgarian Academy of Sciences, 	Acad. G. Bonchev St. 8, Sofia 1113, Bulgaria}

\author{Daniela D. Doneva}
\email{daniela.doneva@uni-tuebingen.de}
\affiliation{Theoretical Astrophysics, Eberhard-Karls University
	of T\"ubingen, T\"ubingen 72076, Germany}
\affiliation{INRNE - Bulgarian Academy of Sciences, 1784  Sofia, Bulgaria}

\author{Kostas~D.~Kokkotas}
\email{kostas.kokkotas@uni-tuebingen.de}
\affiliation{Theoretical Astrophysics, Eberhard-Karls University
	of T\"ubingen, T\"ubingen 72076, Germany}

\begin{abstract}
The recent detection of gravitational waves from  a neutron star merger was a significant step towards constraining the nuclear matter equation of state by using the tidal Love numbers (TLNs) of the merging neutron stars. Measuring or constraining the neutron star TLNs allows us in  principle to exclude or constraint many equations of state. This approach, however, has the drawback that many modified theories of gravity could produce deviations from General Relativity similar to the deviations coming from the uncertainties in the equation of  state. The first and the most natural step in resolving the mentioned problem is to quantify the effects on the TLNs from the modifications of General Relativity. With this motivation in mind, in the present paper we calculate the TLNs of (non-rotating) neutron stars in $f(R)$ gravity. For this purpose, we first derived the equations describing both the polar and the axial stationary perturbations of neutron stars in a particular class of $f(R)$ gravity, the so-called $R^2$-gravity. Then, by solving numerically the perturbation equations, we calculate explicitly the polar and the axial $l=2$ TLNs  of the neutron stars in $R^2$-gravity for three characteristic realistic equations of state. Our results show that while the polar TLNs are slightly influenced by the $R^2$ modification of General Relativity, the axial TLNs can be several times larger (in terms of the absolute value) compared to the general relativistic case.          
\end{abstract}


\maketitle

\section{Introduction}

The first detection of binary neutron star mergers \cite{Abbott2017} contributed in several ways to the efforts of constraining the nuclear matter equation of state (EOS) \cite{Bauswein2017,Annala2017,Margalit2017,Radice2018,Rezzolla2018,Ruiz2018,Shibata2017,Most2018}. Perhaps one of the most elegant and straightforward constraint comes from the measurement of the tidal Love numbers (TLNs) of the merging neutron stars. On its basis one can already exclude a large number of modern equations of state. This approach, though, have the drawback that many alternative theories of gravity would produce deviations from pure general relativity (GR) similar in magnitude and characteristics to the uncertainties in the EOS \footnote{En escape from this problem can be offered for example by the universal, i.e. equation of state independent relations (see e.g. \cite{Yagi2013,Yagi2013a,Doneva2015,Doneva2014a,Kleihaus2014,Breu2016,Staykov2016}).}. Thus, it is difficult to determine the EOS from the current and forthcoming gravitational wave observations in a theory of gravity independent way. The first step in solving this problem is to quantify the effects from modification of general relativity on the TLNs. 

The TLNs characterize the response (deformability) of a body to an external tidal force \cite{Murray1999,Poisson2014}. They encode information about the internal structure of the body and the strong field regime of gravity  and most importantly -- the tidal Love numbers can be determined through the gravitational wave emission of merging neutron stars \cite{Flanagan2008,Hinderer2008,Binnington2009,Damour2009,Hinderer2010,Vines2011,Damour2012,DelPozzo2013,Maselli2013a,Hinderer2016}. They are very scarcely studies in alternative theories of gravity with the only exception of the dynamical Chern-Simons gravity \cite{Yagi2013,Yagi2013a}. The TLNs of different exotic compact objects and for BHs in some alternative theories of gravity (having nonzero TLNs in contrast to pure GR) were examined in \cite{Cardoso2017}.

The interest in different modifications of GR on the other hand is growing in the last decade. The reason steams from the fact that from one hand there are phenomena such as the accelerated expansion of the Universe, that can not be explained well within Einstein's theory of gravity without requiring fine-tuning or other problems appearing. From the other hand there are purely theoretical arguments coming from the theories trying to unify all the interactions, the quantum corrections to the GR Lagrangian, or the attempt to quantize gravity. In this context the $f(R)$ theories  of gravity, where the Ricci scalar in the Einstein-Hilbert action is substituted by a function of this scalar,  received particular interest recently \cite{Sotiriou2010,DeFelice2010a,Nojiri2011}. The reason is that they can serve as one of the promising alternative explanation of the dark energy phenomenon. Also, due to the quantum corrections in the strong field regime, the renormalization at one loop requires that the Einstein-Hilbert action be supplemented with higher order terms similar to $f(R)$ theories \cite{Birrell1984}. 
 
The $f(R)$ theories, though, are explored mainly on cosmological scales because of the connection to the dark energy problem and compact objects in these theories are more scarcely studied. Non-perturbative models of neutron stars were constructed in the static case in these theories in \cite{Babichev2010,Yazadjiev2014,Astashenok2017}  and they were later extended to the slowly \cite{Staykov2014} and rapidly rotating \cite{Yazadjiev2015} cases.  Binary neutron star mergers in $R^2$ gravity were examined in \cite{Sagunski2018}. The goal of the present paper is to calculate the TLN of neutron stars in $f(R)$ theories and more precisely, in a particular class of $f(R)$ theories that is supposed to give the dominant contribution on astrophysical scales, namely the $R^2$-gravity having a Lagrangian of the form $f(R)=R+aR^2$ where $a$ is a free parameter. The current observational constraints impose the following upper bound $a \lesssim 10^{11} \rm{m}^2$ \cite{Naf2010} which leaves space for significant deviations from pure GR. Even though we concentrate on $f(R)$ theories, the general framework for calculating TLN developed in the paper is valid for a much larger class of alternative theories of gravity, the massive scalar-tensor theories. The reason comes from the fact that $f(R)$ theories are mathematically equivalent to a particular class of scalar-tensor theory with nonzero scalar field potential. Even more, we use this equivalence explicitly in the present paper in order to simplify the calculations.

The paper is organized as follows. The mathematical framework behind the $R^2$-gravity and the way of constructing equilibrium neutron star solutions is examined in Section I. In Section II the formulas for the calculation of the TLN, both polar and axial, are derived. Section III is devoted on the numerical results and the comparison with pure GR. The paper ends with Conclusions.

\section{f(R) theories and equivalence to scalar-tensor theories}
The essence of $f(R)$ theories is that the Ricci scalar $R$ in the action is substituted by a function of this scalar $f(R)$:
\begin{eqnarray}\label{eq:action_FR}
	S= \frac{1}{16\pi G} \int d^4x \sqrt{-g} f(R) + S_{\rm
		matter}(g_{\mu\nu}, \chi).
\end{eqnarray}
Here $R$ is the Ricci scalar with respect to the spacetime metric $g_{\mu\nu}$ and $S_{\rm matter}$ is the action of the matter where the matter fields are denoted by $\chi$. If we want the theory to be well posed, i.e. to be free of tachyonic instabilities and ghosts, the following inequalities should be fulfilled  $d^2f/dR^2\ge 0$ and $df/dR>0$. We will work in a particular class of $f(R)$ theories, the so-called $R^2$ gravity where
\begin{equation}
f(R) = R + aR^2,
\end{equation}
which is supposed to give the dominant contribution for strong fields. Here $a$ is a parameter and in order to satisfy the above given inequalities one should require that $a\ge 0$.

A very common approach is to work not with the original form of the action but transform it to another one by substituting $\Phi=\frac{df(R)}{dR}$ and
$U(\Phi)=R \frac{df}{dR} - f(R)$. In this way we obtain an action that is equivalent to a particular class of the  Brans-Dicke theory with a parameter $\omega_{BD}=0$:
\begin{eqnarray}\label{eq:STT_JF}
	S=\frac{1}{16\pi G} \int d^4x \sqrt{-g}\left[\Phi R - U(\Phi)\right]
	+ S_{\rm matter}(g_{\mu\nu}, \chi).
\end{eqnarray}
In the case of $R^2$ gravity the potential takes the form 
\begin{eqnarray}
U(\Phi)=\frac{1}{4a}(\Phi - 1)^2,
\end{eqnarray}
and therefore the scalar field is massive with 
\begin{equation}\label{eq:mass_scalar_field}
m_{\Phi}=\frac{1}{\sqrt{6a}}. 
\end{equation}

It is mathematically equivalent to work either with the original form of the action \eqref{eq:action_FR} or with its scalar-tensor representation \eqref{eq:STT_JF} and this was explicitly shown also in \cite{Yazadjiev:2015xsj} for the case of neutron star solutions.

The action \eqref{eq:STT_JF} is written in the physical Jordan frame where there is no direct coupling between the matter and the scalar field in order to satisfy the weak equivalence principle. One can further simplify the problem by introducing the Einstein frame by making a conformal transformation of the metric 
\begin{equation}
g^{*}_{\mu\nu}=\Phi g_{\mu\nu}
\end{equation}
and redefining the scalar field and the potential
\begin{equation}
\varphi =\frac{\sqrt{3}}{2}\ln\Phi,\;\;\;  V=\frac{U(\Phi)}{\Phi^2}.
\end{equation}
Thus we arrive at the following Einstein frame action
\begin{eqnarray}\label{EFA}
	S=\frac{1}{16\pi G} \int d^4x \sqrt{-g^{*}}\left[ R^{*} - 2
	g^{*\mu\nu}\partial_{\mu}\varphi \partial_{\nu}\varphi - V(\varphi)
	\right] + S_{\rm
		matter}(A^2(\varphi)g^{*}_{\mu\nu},\chi),
\end{eqnarray}
where $R^{*}$ is the Ricci scalar curvature with respect to the Einstein frame metric $g^{*}_{\mu\nu}$. As one can see, a direct couping between the matter and the scalar field appears in the Einstein frame through the coupling function $A^2(\varphi)=\Phi^{-1}(\varphi)$. In the particular case of  $R^2$ gravity the coupling function and the scalar field potential take the following form
\begin{equation}
A(\varphi)=e^{-\frac{1}{\sqrt{3}}\varphi}, \;\;\; V(\varphi)= \frac{1}{4a}
\left(1-e^{-\frac{2\varphi}{\sqrt{3}}}\right)^2.
\end{equation}

The field equations in the Einstein frame are much simpler compared to the Jordan frame ones and that is why we will employ this frame. Of course, the final quantities that we obtain have to be transformed back to the physical Jordan frame. In addition, the equation of state of the nuclear matter that we use will be also only in the Jordan frame.  A more detailed discussion  of the problem can be found in \cite{Yazadjiev2014,Staykov2014,Yazadjiev2015}.

We will consider nonrotating stars and thus the following general ansatz for the static and spherically symmetric Einstein frame metric can be used
\begin{eqnarray}
&&ds_{*}^2 = -e^{2\psi} dt^2 + e^{2\Lambda} dr^2 + r^2(d\theta^2+ \sin^2\theta\; d\phi^2),
\end{eqnarray}
where all the metric functions depend on $r$ only. The reduced field equations take the following form

\begin{eqnarray}
&&\frac{1}{r^2}\frac{d}{dr}\left[r(1- e^{-2\Lambda})\right]= 8\pi G
A^4(\varphi) \rho + e^{-2\Lambda}\left(\frac{d\varphi}{dr}\right)^2
+ \frac{1}{2} V(\varphi), \label{eq:FieldEq1} \\
&&\frac{2}{r}e^{-2\Lambda} \frac{d\psi}{dr} - \frac{1}{r^2}(1-
e^{-2\Lambda})= 8\pi G A^4(\varphi) p +
e^{-2\Lambda}\left(\frac{d\varphi}{dr}\right)^2 - \frac{1}{2}
V(\varphi),\label{eq:FieldEq2}\\
&&\frac{d^2\varphi}{dr^2} + \left(\frac{d\psi}{dr} -
\frac{d\Lambda}{dr} + \frac{2}{r} \right)\frac{d\varphi}{dr}= 4\pi G
\alpha(\varphi)A^4(\varphi)(\rho-3p)e^{2\Lambda} + \frac{1}{4}
\frac{dV(\varphi)}{d\varphi} e^{2\Lambda}, \label{eq:FieldEq3}\\
&&\frac{dp}{dr}= - (\rho + p) \left(\frac{d\psi}{dr} +
\alpha(\varphi)\frac{d\varphi}{dr} \right). \label{eq:FieldEq4}
\end{eqnarray}
Here the Jordan frame pressure $p$ and energy density $\rho$ are used and they are connected to the Einstein frame ones ($p_*$ and $\rho_*$) via the following relations  $p^*=A^{4}(\varphi)p$ and $\rho^*=A^{4}(\varphi) \rho$. The Jordan frame quantities $\rho$ and $p$ are naturally connected via the equation of state (EOS) for the neutron star matter $p=p(\rho)$. In addition, we have to impose the standard boundary conditions -- regularity at the center of the star and asymptotic flatness at infinity. 

The radius of the star is calculated from the requirement of vanishing of the pressure at the stellar surface and the mass is taken from the asymptotic expansion of the metric functions at infinity. It is important to note that for the considered $R^2$ gravity the mass in the Einstein and the Jordan frame coincide, while the physical Jordan frame radius of the star $R_S$ is connected to the Einstein frame one $r_s$ in the following way
\begin{eqnarray}
R_{S}= A[\varphi(r_S)] r_S.
\end{eqnarray}

Here we have presented the problem of calculating the background equilibrium neutron star solutions very briefly. More detailed explanations can be found in \cite{Yazadjiev2014}.

In what follows, we shall use the dimensionless parameter $a\to a/R^2_{0}$, where $R_{0}$ is one half of the solar gravitational radius   $R_{0}=1.47664 \,{\rm km}$.

\section{Tidal Love numbers}

In order to compute the tidal Love numbers we have to consider the stationary  perturbations of the static and spherically symmetric stars in $R^2$-gravity . The perturbations of the metric can be separated in polar and axial type. Here we present the two cases separately and derive the tidal Love numbers in both cases.

\subsection{Polar}

For the polar perturbations the peturbed Einstein frame metric in the Regger-Wheeler gauge can be written in the form

\begin{eqnarray}
H^{polar}_{\mu\nu} = \left(
\begin{array}{cccc}
-e^{2\psi_{0}}H_{0}(r) & H_{1}(r) & 0 & 0 \\
H_{1}(r) & e^{2\Lambda_{0}} H_{2}(r) & 0 & 0 \\
0 & 0 &   K(r) r^2 & 0\\
0 & 0 & 0 & K(r) r^2\sin^2\theta  \\
\end{array} \right) Y_{lm}(\theta,\phi), 
\end{eqnarray}   
where $Y_{lm}(\theta,\phi)$ are the spherical harmonics. The perturbations of the scalar field, energy density and the pressure can be decompose in the form
$\delta\varphi=\delta\tilde\varphi(r) Y_{lm}(\theta,\phi)$, $\delta\rho^*=\delta\tilde\rho(r) Y_{lm}(\theta,\phi)$ and $\delta p^*=\delta\tilde p(r) Y_{lm}(\theta,\phi)$. After perturbing the Einstein frame field equations of the $f(R)$ gravity coupled  to a perfect fluid  it can be shown that $H_0=-H_2$ and  $H_1=0$. Also $K$, $\delta\tilde\rho(r)$ and $\delta\tilde p(r)$ can be written as functiond of $H_{0}$ and $\delta\tilde \varphi$. Finally we obtain two equations for $H_0=-H_2=H$ and $\delta\tilde \varphi$ governing the stationary perturbations of the static and sphereically symmetric stars
in $f(R)$ gravity:
\begin{eqnarray}
\frac{d^2H}{dr^2} &+& \left\{ \frac{2}{r} + e^{2\Lambda_0} \left[ \frac{1-e^{-2\Lambda_0}}{r} + 4\pi(p^*_{0}-\rho^*_0)r - \frac{1}{2}
V(\varphi_0)r \right] \right\} \frac{dH}{dr} \notag \\
&+& \left\{ -\frac{l(l+1)}{r^2}e^{2\Lambda_0} + 4\pi e^{2\Lambda_0}\left[9p^*_0 + 5\rho^*_0  + \frac{\rho^*_0+p^*_0}{\tilde{c}_s^2} -\frac{1}{4\pi} V(\varphi_0) \right] - 4 \psi_0'^2  \right\} H \notag \\
\notag \\
&+& e^{2\Lambda_0} \Big\{ -4 \varphi_0' r \left[ \frac{1-e^{-2\Lambda_0}}{r^2} +8\pi p^*_0 + e^{-2\Lambda_0} \varphi_0'^2 - \frac{1}{2}
V(\varphi_0) \right] \notag \\
&&\;\;\;\;\;\;\;\;\; -   \frac{16\pi}{\sqrt{3}}\left[ (\rho^*_0-3p^*_0) + (\rho^*_0+p^*_0)\frac{1-3\tilde{c}_s^2}{2\tilde{c}_s^2} \right]  + V^\prime(\varphi_{0})\Big\}\delta\tilde{\varphi}=0 \label{eq:PertEq_H_polar} \\  \notag \\  \notag \\
\frac{d^2 \delta\tilde{\varphi}}{dr^2} &+& \left(\psi_0' - \Lambda_0' +\frac{2}{r}\right) \frac{d\delta\tilde{\varphi}}{dr} \notag \\
&-&e^{2\Lambda_0}\Big\{ \frac{l(l+1)}{r^2} + 4 e^{-2\Lambda_0} \varphi_0'^2 + \frac{1}{4}V''(\varphi_0)  
-  \frac{8\pi}{3} \left[-2(\rho^*_0-3p^*_0) + (\rho^*_0+p^*_0)\frac{1-3\tilde{c}_s^2}{2\tilde{c}_s^2}\right] \Big\}  \delta\tilde{\varphi} \notag \\
&+& e^{2\Lambda_0}\left\{ -2 e^{-2\Lambda_0}\psi_0'\varphi_0' - \frac{4\pi}{\sqrt{3}} \left[(\rho^*_0-3p^*_0) + (\rho^*_0+p^*_0)\frac{1-3\tilde{c}_s^2}{2\tilde{c}_s^2}\right] + \frac{1}{4} V'(\varphi_0) \right\} H = 0 \label{eq:PertEq_phi_polar}
\end{eqnarray}
Here $\Lambda_0$, $\psi_0$, $\varphi_0$, $p^*_0$ and $\rho^*_0$ are the corresponding unperturbed variables taken from the background neutron star solutions and $\tilde c_s$ is the Jordan frame sound speed defined by  ${\tilde c_s}^2= \partial p/\partial \rho$. 

In the considered model the scalar field mass is nonzero which means that both the background scalar field $\varphi_0$ and its perturbation $\delta\tilde{\varphi}$ drop off exponentially at infinity. This means that the corresponding scalar field tidal Love number is zero.

The asymptotic behavior of $H$ at large $r$  on the other hand  is
\begin{equation}
H=\frac{c_1}{r^{l+1}} + {\cal O}\left(\frac{1}{r^{l+2}}\right) + c_2 r^l + {\cal O}(r^{l-1}).
\end{equation}
The tidal Love number $k_2$ is connected to the coefficients in the above given expansion $c_1$ and $c_2$ in the following way:
\begin{equation}
k_l^{polar} = \frac{1}{2R_{S}^{2l+1}} \frac{c_1}{c_2}.
\end{equation}

In pure GR the ratio $ c_1/c_2$ is usually determined after matching at the stellar surface the numerical solution for $H$ inside the neutron star to the analytical solution outside it \cite{Flanagan2008,Hinderer2008,Binnington2009,Damour2009}. Applying this approach directly to our problem is not possible since the equations for $H$ and $\delta\tilde{\varphi}$ are coupled and in the general case no analytical solution exist outside the star. The fact that the scalar field is massive, though, simplifies the problem a lot. Since both the scalar field and its perturbation die out exponentially at  distances larger than the Compton wavelength of the scalar field $\lambda_\varphi=2\pi/m_\varphi$, far enough from the surface of the star the  scalar field and its perturbation are practically zero and we can use the same analytical solution as in pure GR. This requires, of course, a matching of the inner numerical and outer analytical solutions to be done not at the surface of the star but far enough from the stellar surface where the scalar field and its perturbation are negligible. This large distance where we match the two solutions will be denoted by $r_{\rm match}$. We have also verified that matching the two solutions not at the stellar surface but at $r_{\rm match}$,  works very well and does not lead to any numerical problems.

The perturbation equation for $H$ far away from the center of the star, where the scalar field and its perturbation are negligible, can be obtained straightforward  from eq. \eqref{eq:PertEq_H_polar} after substituting $p^*_0=\rho^*_0=\varphi_0=\delta\tilde{\varphi}=0$. This equation is the same as in pure GR and its analytical solution can be expressed in terms of elementary functions \cite{Hinderer2008}. As commented, the value of $k_2$ can be calculated after matching the numerical and analytical solutions at large enough radial distances $r_{\rm match}$ and for the $l=2$ case one obtains\footnote{Since $r_{\rm match}$ is connected to the Compton wavelength of the scalar field, $r_{\rm match}$ is not a constant but increases with the increase of the parameter $a$.}
\begin{eqnarray}
k_2^{polar} &=& \frac{8C_1^5}{5}(1-2C)^2\left[2+2C(y-1)-y\right]\times \notag \\
&&\Big\{ 2C(6-3y+3C(5y-8))+4C^3\left[13-11y+C(3y-2) + 2C^2(1+y)\right] \notag \\
&&\;\; + 3(1-2C)^2\left[2-y+2C(y-1)\right]\log(1-2C)\Big\}^{-1}
\end{eqnarray}
where $y=rH'/H$, $C_1=M/R_{S}$ is the compactness of the star and $C=M/r_{\rm match}$. The value of $y$ is calculated after solving numerically the coupled system of equations \eqref{eq:PertEq_H_polar}, \eqref{eq:PertEq_phi_polar} from $r=0$ to $r=r_{\rm match}$.

It is important to note, that the polar TLNs are the same in the physical Jordan and the Einstein frame since the scalar field drops off exponentially outside the star. 

\subsection{Axial}

In the axial case the metric perturbations are given by 

\begin{eqnarray}
H^{axial}_{\mu\nu} = \left(
\begin{array}{cccc}
0 & 0 &  h(r) S^{lm}_{\theta}(\theta,\phi) & h(r)S^{lm}_{\phi}(\theta,\phi) \\
0 & 0 &  h_{1}(r)S^{lm}_{\theta}(\theta,\phi)& h_{1}(r)S^{lm}_{\phi}(\theta,\phi) \\
h(r)S^{lm}_{\theta}(\theta,\phi) & h_1(r)S^{lm}_{\theta}(\theta,\phi) & 0 & 0 \\
h(r)S^{lm}_{\phi}(\theta,\phi) & h_1(r)S^{lm}_{\phi}(\theta,\phi)  & 0 & 0  \\
\end{array}
\right), 
\end{eqnarray}
where $(S^{lm}_{\theta}(\theta,\phi),S^{lm}_{\phi}(\theta,\phi))=(-\partial_{\phi}Y_{lm}(\theta,\phi)/\sin\theta,\,\sin\theta \partial_{\theta}Y_{lm}(\theta,\phi))$. 
The perturbations of the scalar field, the energy density and the pressure 
vanish. Using the perturbations of the field equations one can show that $h_1=0$.   
We are left with only one equation for the metric perturbation $h$:
\begin{eqnarray}
\frac{d^2h}{dr^2} &-&\left[ 4\pi (\rho^*_{0} + p^*_{0})e^{2\Lambda_{0}} + \left(\frac{d\varphi_{0}}{dr}\right)^2  \right]r \frac{dh}{dr} \nonumber \\
&+& \left[- \frac{(l-1)(l+2)}{r^2}e^{2\Lambda_{0}}  + 8\pi (\rho^*_{0} + p^*_{0})e^{2\Lambda_{0}} + 2 \left(\frac{d\varphi_{0}}{dr}\right)^2  -\frac{2}{r^2} \right] h= 0 
\end{eqnarray}
In this case we also do not have an analytical solution because of the presence of scalar field terms. Similar to the polar case, though, such solution can be found far outside the star where the scalar field has died out exponentially and the solution is the same as in pure GR. The asymptotic equation at such large distances is 
\begin{eqnarray}\label{eq:PertEq_h_axial}
\left(1-\frac{2M}{r}\right) \frac{d^2h}{dr^2} + \left[- \frac{l(l+1)}{r^2} + \frac{4M}{r^3} \right]h=0
\end{eqnarray}
and it can be solved analytically for a given $l$ \cite{Damour2009,Binnington2009}. The function $h$ has the following asymptotic:
\begin{eqnarray}
h \approx \frac{c_1}{r^l} + c_{2}r^{l+1} 
\end{eqnarray}
The tidal Love numbers for the axial perturbations $k^{axial}_{l}$ are related to the coefficients $c_1$ and $c_2$ in the following way
\begin{eqnarray}
k^{axial}_{l}= - \frac{l}{2(l+1)} \frac{c_1}{c_2} \frac{1}{ R_{S}^{2l+1}}.
\end{eqnarray}

The value of $k^{axial}_{l}$ can be found after matching the numerical solution (obtained after integrating the perturbation equation \eqref{eq:PertEq_h_axial} from the center of the star to some large distance $r_{\rm match}$ were the scalar field is negligible) with the analytical asymptotic solution. Thus, one can obtain the following relation for the $l=2$ case \footnote{The definition of $k^{axial}_{2}$ differs from the one in \cite{Damour2009} by a constant factor of 12.}
\begin{eqnarray}
k^{axial}_{2}= - \frac{8 C_1^5}{5} \frac{2C (y-2) - y + 3}{2C \left[2C^3(y+1) + 2C^2 y + 3C(y-1) - 3y+9\right] + 3\left[2C(y-2) - y + 3\right]\log(1-2C)}
\end{eqnarray}
where $y=rh'/h$, $C_1=M/R_{S}$ is the compactness of the star and $C=M/r_{\rm match}$.

Similar to the polar case, the axial TLNs are the same in the physical Jordan and the Einstein frame.

\section{Numerical results}
\begin{figure}[]
	\centering
	\includegraphics[width=0.45\textwidth]{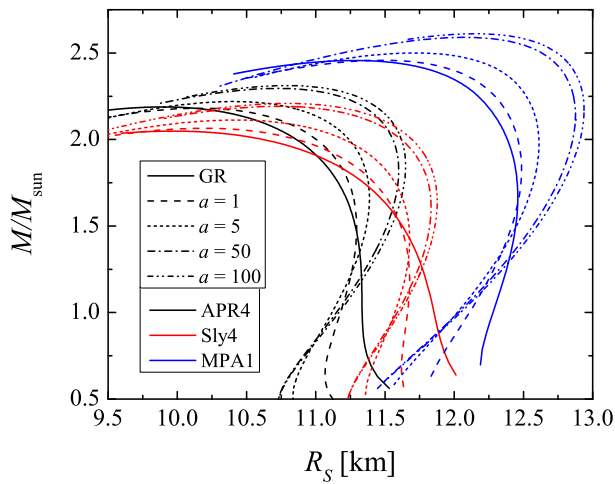}
	\caption{The mass as a function of the radius for all three considered EOSs and different values of the parameter $a$, including the pure GR case. }
	\label{Fig:M_R}
\end{figure}

We will work with three modern realistic EOS that allow for models with maximum mass above the two solar mass barrier \cite{Antoniadis13,Demorest10} and are in agreement with the constraints coming from the observation of the tidal Love numbers of merging neutron stars \cite{Abbott2017}. These are the APR4 EOS \cite{AkmalPR}, the SLy4 EOS \cite{Douchin2001} and the MPA1 EOS \cite{Muether1987}. We should note, though, that the MPA1 EOS actually do not fit well in the current constrains coming from the electromagnetic observations \cite{Lattimer2014,Oezel2016} but we included it in our studies so that we can cover a larger range of stiffness.  In order to be able to make a better comparison between the three EOS and judge about the effect of $R^2$ gravity on the background neutron star models, the mass as a function of the radius is plotted in Fig. \ref{Fig:M_R} for these EOSs and for four indicative values of the parameter $a$. The chosen values of $a$ are the same as the ones used in the calculations of the tidal Love numbers. Since $a$ is back-proportional to the mass of the scalar field (see eq. \eqref{eq:mass_scalar_field}), $a\rightarrow\infty$ corresponds to the massless scalar field case. 
On the other hand the mass of the scalar field goes to infinity when $a\rightarrow 0$ which means that the Compton wavelength is practically zero and the corresponding solutions tend to the pure GR case. As one can see in Fig. \ref{Fig:M_R} the differences of the neutron star masses and radii with pure GR can reach up to roughly 10\% and they are comparable both qualitatively and quantitatively with the deviations due to the EOS uncertainties.  

Let us comment in more detail the particular values of $a$ that we have chosen. More precisely, we have worked with $a \le 100$ because of the following reason. The parameter $a$ introduces a length-scale related to the Compton wavelength of the scalar field, above which the scalar field drops off exponentially and thus the scalar field has a finite range \footnote{In terms of the original $f(R)$ formulation of the theory, this is the range beyond which the $R^2$ type of modifications of general relativity are negligible.}. This is a crucial ingredient in our calculations of the TLNs since we use the pure GR solution of the perturbation equations outside this Compton radius. Thus, it is natural to  require that the Compton wavelength of the scalar field is smaller than the orbital separation between the merging neutron stars at the time when they can be observed by the ground base detectors and the TLNs can be measured \footnote{Calculating the TLNs when the orbital separation is smaller than the Compton wavelength or the scalar field is not massive, is much more complicated and was not solved until now in any alternative theory of gravity. We are currently working on this problem.}. 
Assuming that with the current instruments we can detect the emitted gravitational waves when the orbital separation drops down to roughly a few hundreds of kilometers, we have chosen to work with $a\le 100$ which leads to $\lambda_\varphi \le 226{\rm km}$.

\begin{figure}[]
	\centering
	\includegraphics[width=0.45\textwidth]{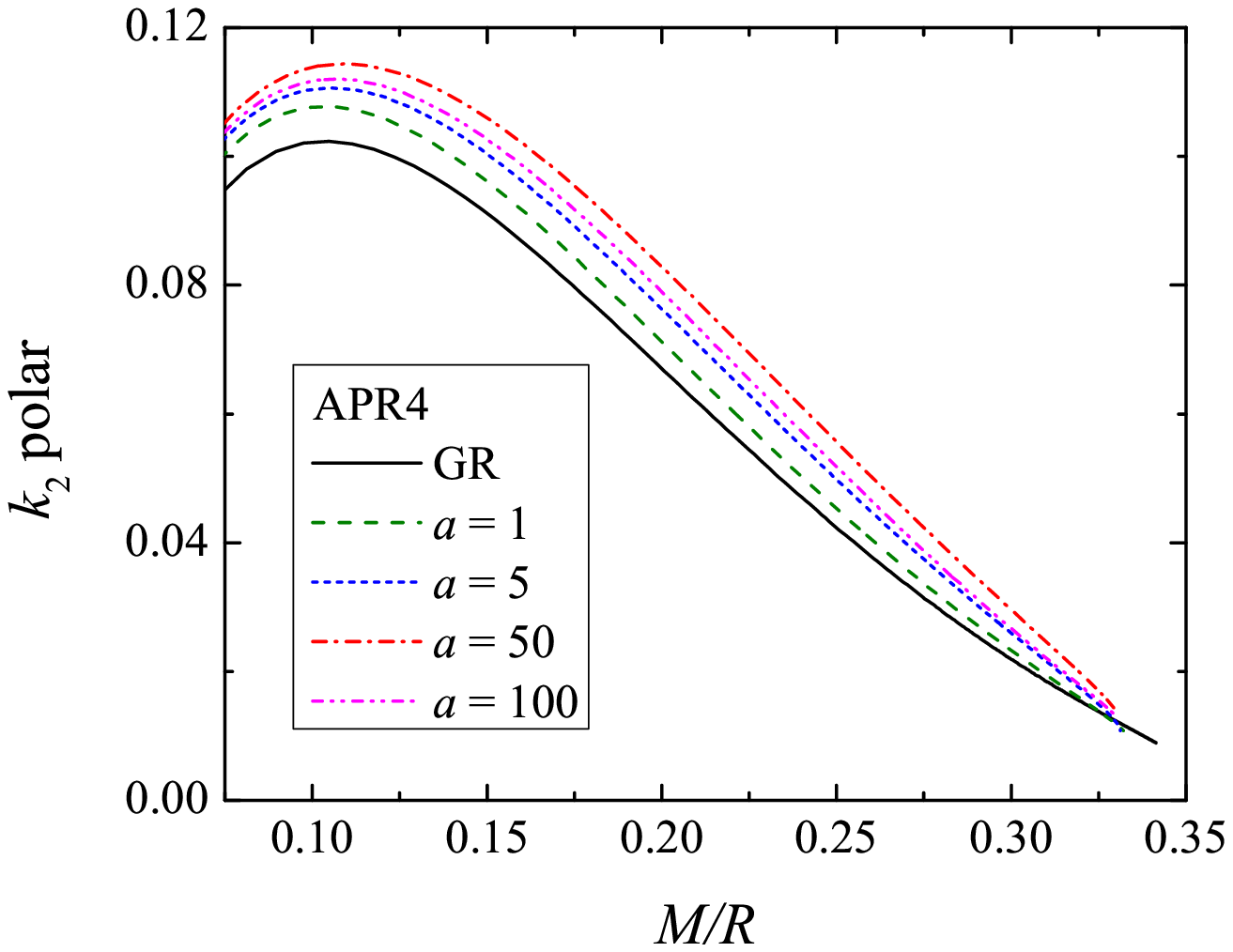}
	\includegraphics[width=0.45\textwidth]{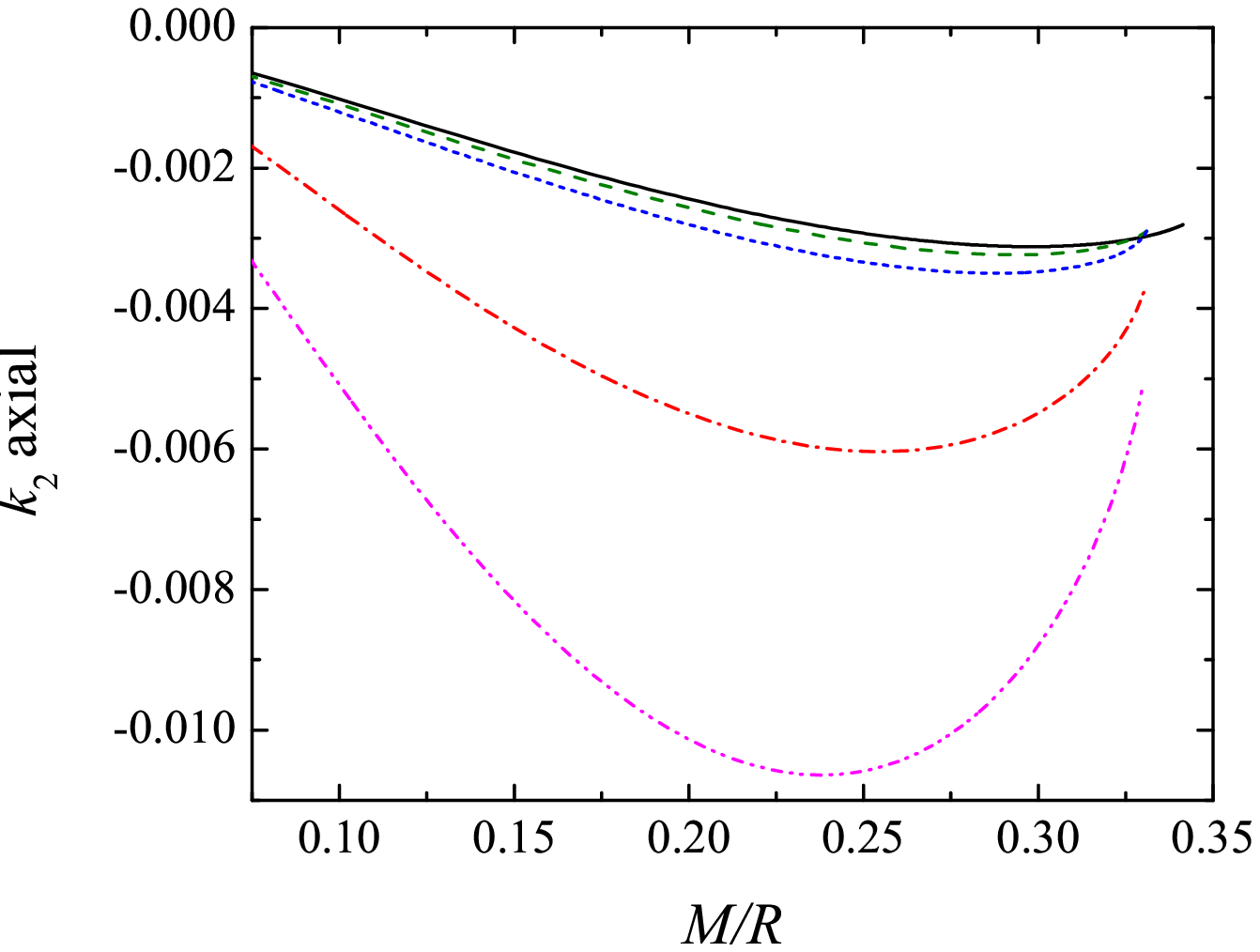}
	\caption{The polar (left panel) and axial (right panel) tidal Love numbers as functions of the stellar compactness for the ARP4 EOS and several values of the parameters $a$.}
	\label{Fig:k2_M}
\end{figure}

\begin{figure}[]
	\centering
	\includegraphics[width=0.45\textwidth]{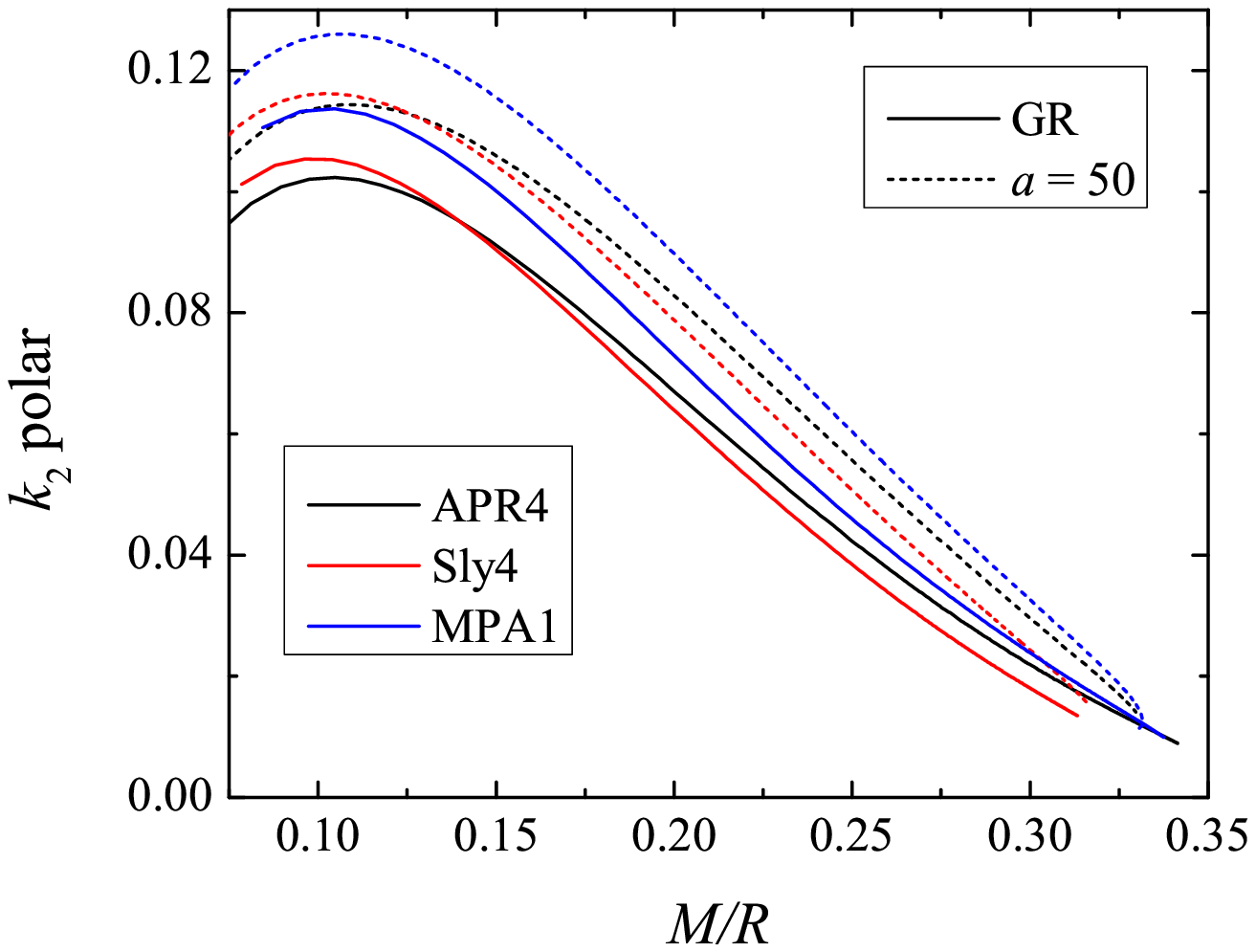}
	\includegraphics[width=0.45\textwidth]{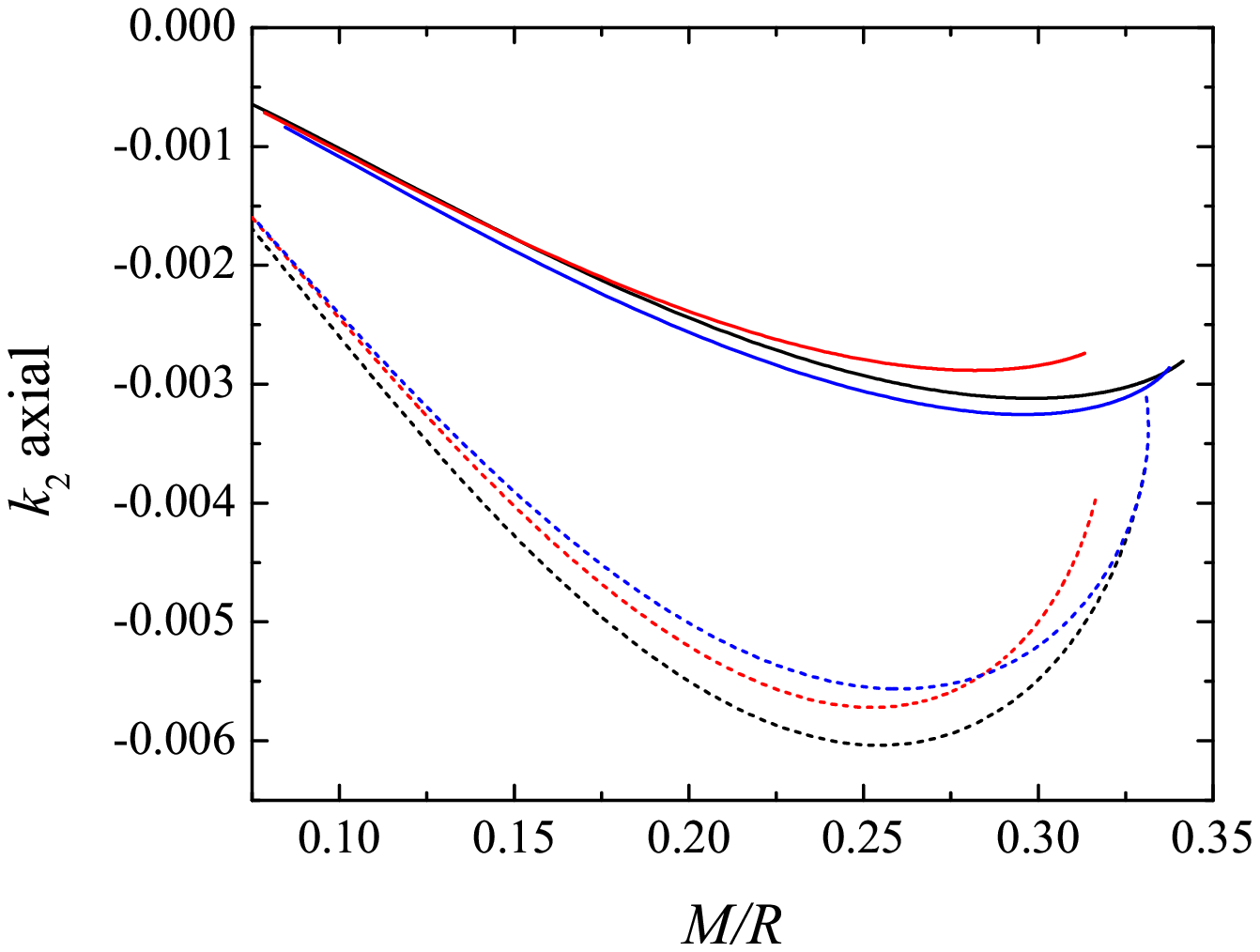}
	\caption{The polar (left panel) and axial (right panel) tidal Love numbers as functions of the stellar compactness for several EOS, and for the pure GR case and $a=50$.}
	\label{Fig:k2_EOS}
\end{figure}

The polar (left panel) and the axial (right panel) TLNs as functions of the neutron star compactness are plotted in Fig. \ref{Fig:k2_M} for four values of $a$ and only for the APR4 EOS, in order to have better visibility. The equation of state dependence of the results is presented in Fig. \ref{Fig:k2_EOS}. As one can see for a fixed equation of state the polar tidal Love number can vary up to roughly 10\% for the considered range of values of $a$ while the variation in the axial Love number is much larger -- for $a=100$ it can differ approximately 4 times. These deviations from the pure GR case can be larger for larger values of $a$, but as we commented, we have limited our studies to $a\le100$.

Clearly, the changes in the polar Love numbers due to $R^2$-gravity are within the current equation of state uncertainty. Hence, the current generation of gravitational wave detectors is unlikely to be able to set constraints on the parameter $a$. The three EOSs, though, were chosen to be the ones allowed by the measurement of the TLNs in the neutron star merger \cite{Abbott2017}. If we take into account also the constraints from the electromagnetic observations \cite{Lattimer2014,Oezel2016} the picture might change because then the MPA1 EOS is outside the allowed range of masses and radii. Thus, if we consider only the APR4 and Sly4 EOSs, the maximum deviation in $R^2$-gravity is larger than the difference between the two equations of state. A definite answer whether this is an observable effect or not can be given only after a detailed analysis of change in the phase of the signal and such a study is underway. 

On the other hand the electromagnetic observations are rapidly advancing and the next generation of gravitational wave detectors is already planned. That is why one can expect that when we know the EOS with a better accuracy in the future from the electromagnetic observations, and have more accurate observations of the gravitational waveforms of merging neutron stars, the $R^2$-gravity effect will be important, producing effects larger than the equation of state uncertainties. Thus we  should take them into account when extracting the relevant parameters from the gravitational wave signal.

The axial TLNs, on the other hand, are significantly influenced (for $a\le100$) by the modifications of the theory of gravity. Moreover, the absolute value of the axial tidal Love numbers increase compared to the pure GR case. In pure GR the contribution of the axial TLN to the gravitational wave phase is one-two orders of magnitude smaller that the polar contribution and the contribution of the higher order (higher $l$) polar TLN would be more important than the axial one as far as the change of the phase in the gravitational wave signal is concerned \cite{Damour2009}. Having axial TLN in $R^2$-gravity that are 4 times larger in terms of absolute values would bring the axial contribution at least of the same order as the higher $l$ polar contribution. With the current accuracy of the detectors only the leading $l=2$ polar TLN could be measured but the axial TLN in $R^2$-gravity might be important for the next generation of gravitational wave detectors. This can potentially be used to constraint the deviations from pure GR.

\section{Conclusion}
In the present paper we have calculated the tidal Love numbers of neutron stars in a particular class of $f(R)$ theories, that is supposed to give the dominant contribution for strong fields, namely the $R^2$-gravity. The study is motivated by the fact that the recent detection of gravitational waves from merging neutron stars allowed us to measure their TLN and thus set constraints on the EOS. Since very often there is a degeneracy between effects coming from modifications of GR and uncertainties of the nuclear matter EOS, an estimation of the influence of alternative theories of gravity on the TLN is very important for the proper interpretation of the observational data. As a matter of fact the problem of calculating the TLN in modified gravity is not well studied at all, with the only exception of the dynamical Chern-Simons gravity.

We have calculated both the axial and the polar TLN for $l=2$. The results show that while the polar TLN is only slightly influenced by the modification of GR (the deviations are within the EOS uncertainty is we consider a broader set of EOSs), the axial TLN can be several times larger (in terms of absolute value) compared to pure GR. Of course these conclusions are for values of the free parameter of the $R^2$-gravity that are in agreement with the observations and also fulfill the requirement that the effective radius of action of the $R^2$ term in the Lagrangian is smaller than the orbital separation between the merging neutron stars when they enter inside the detector sensitivity.

The question is whether such deviations from Einstein's theory can lead to observable effects. The problem is that the quantity that can actually be measured with the current detectors is the polar TLN, while the axial one is expected to give much smaller (yet unmeasurable) contribution in the change of the phase of the signal. Taking into account that the absolute value of the axial TLN can increase a lot in $R^2$-gravity, it could have a stronger influence on the signal that can potentially be detected by the next generation of gravitational wave detectors. This would allow us to set constraints on $f(R)$ theories of gravity and to test the strong field regime of GR. On the other hand, when the nuclear matter EOS is better constrained by the electromagnetic observations in the future, the deviations in the polar TLN would also become observationally important in order to be able to accurately interpreted the detected signal. 

As a matter of fact even though the problem of nonzero scalar field mass is more complicated in terms of generating the background neutron star solutions, finding the tidal Love numbers in the massless case is much more involved compared to the massive one. The reason is that the scalar field decreases as $1/r$ at infinity in the massless case, compared to an exponential decay in the massive one, which leads to nonzero dilaton charge and the inability to use the analytic general relativistic exact solutions far away from the source. Exact solutions in the more general case of scalar-tensor theories are currently not available and this would be the first step towards solving the problem in the massless case. Additional complication comes from the fact that we would have a nonzero scalar tidal Love numbers. We are currently working on this problem.

\acknowledgments{DD would like to thank the European Social Fund, the Ministry of Science, Research and the Arts Baden-W\"urttemberg for the support. DD is indebted to the Baden-Württemberg Stiftung for the financial support of this research project by the Eliteprogramme for Postdocs. The support by the COST Actions MP1304, CA15117, CA16104 and CA16214  is also gratefully acknowledged.}

\appendix

\bibliography{references}

\begin{thebibliography}{48}
\expandafter\ifx\csname natexlab\endcsname\relax\def\natexlab#1{#1}\fi
\expandafter\ifx\csname bibnamefont\endcsname\relax
  \def\bibnamefont#1{#1}\fi
\expandafter\ifx\csname bibfnamefont\endcsname\relax
  \def\bibfnamefont#1{#1}\fi
\expandafter\ifx\csname citenamefont\endcsname\relax
  \def\citenamefont#1{#1}\fi
\expandafter\ifx\csname url\endcsname\relax
  \def\url#1{\texttt{#1}}\fi
\expandafter\ifx\csname urlprefix\endcsname\relax\def\urlprefix{URL }\fi
\providecommand{\bibinfo}[2]{#2}
\providecommand{\eprint}[2][]{\url{#2}}

\bibitem[{\citenamefont{Abbott et~al.}(2017)}]{Abbott2017}
\bibinfo{author}{\bibfnamefont{B.}~\bibnamefont{Abbott}} \bibnamefont{et~al.}
  (\bibinfo{collaboration}{Virgo, LIGO Scientific}), \bibinfo{journal}{Phys.
  Rev. Lett.} \textbf{\bibinfo{volume}{119}}, \bibinfo{pages}{161101}
  (\bibinfo{year}{2017}), \eprint{1710.05832}.

\bibitem[{\citenamefont{Bauswein et~al.}(2017)\citenamefont{Bauswein, Just,
  Janka, and Stergioulas}}]{Bauswein2017}
\bibinfo{author}{\bibfnamefont{A.}~\bibnamefont{Bauswein}},
  \bibinfo{author}{\bibfnamefont{O.}~\bibnamefont{Just}},
  \bibinfo{author}{\bibfnamefont{H.-T.} \bibnamefont{Janka}}, \bibnamefont{and}
  \bibinfo{author}{\bibfnamefont{N.}~\bibnamefont{Stergioulas}},
  \bibinfo{journal}{Astrophys. J.} \textbf{\bibinfo{volume}{850}},
  \bibinfo{pages}{L34} (\bibinfo{year}{2017}), \eprint{1710.06843}.

\bibitem[{\citenamefont{Annala et~al.}(2017)\citenamefont{Annala, Gorda,
  Kurkela, and Vuorinen}}]{Annala2017}
\bibinfo{author}{\bibfnamefont{E.}~\bibnamefont{Annala}},
  \bibinfo{author}{\bibfnamefont{T.}~\bibnamefont{Gorda}},
  \bibinfo{author}{\bibfnamefont{A.}~\bibnamefont{Kurkela}}, \bibnamefont{and}
  \bibinfo{author}{\bibfnamefont{A.}~\bibnamefont{Vuorinen}}
  (\bibinfo{year}{2017}), \eprint{1711.02644}.

\bibitem[{\citenamefont{Margalit and Metzger}(2017)}]{Margalit2017}
\bibinfo{author}{\bibfnamefont{B.}~\bibnamefont{Margalit}} \bibnamefont{and}
  \bibinfo{author}{\bibfnamefont{B.~D.} \bibnamefont{Metzger}},
  \bibinfo{journal}{Astrophys. J.} \textbf{\bibinfo{volume}{850}},
  \bibinfo{pages}{L19} (\bibinfo{year}{2017}), \eprint{1710.05938}.

\bibitem[{\citenamefont{Radice et~al.}(2018)\citenamefont{Radice, Perego,
  Zappa, and Bernuzzi}}]{Radice2018}
\bibinfo{author}{\bibfnamefont{D.}~\bibnamefont{Radice}},
  \bibinfo{author}{\bibfnamefont{A.}~\bibnamefont{Perego}},
  \bibinfo{author}{\bibfnamefont{F.}~\bibnamefont{Zappa}}, \bibnamefont{and}
  \bibinfo{author}{\bibfnamefont{S.}~\bibnamefont{Bernuzzi}},
  \bibinfo{journal}{Astrophys. J.} \textbf{\bibinfo{volume}{852}},
  \bibinfo{pages}{L29} (\bibinfo{year}{2018}), \eprint{1711.03647}.

\bibitem[{\citenamefont{Rezzolla et~al.}(2018)\citenamefont{Rezzolla, Most, and
  Weih}}]{Rezzolla2018}
\bibinfo{author}{\bibfnamefont{L.}~\bibnamefont{Rezzolla}},
  \bibinfo{author}{\bibfnamefont{E.~R.} \bibnamefont{Most}}, \bibnamefont{and}
  \bibinfo{author}{\bibfnamefont{L.~R.} \bibnamefont{Weih}},
  \bibinfo{journal}{Astrophys. J.} \textbf{\bibinfo{volume}{852}},
  \bibinfo{pages}{L25} (\bibinfo{year}{2018}), \eprint{1711.00314}.

\bibitem[{\citenamefont{Ruiz et~al.}(2018)\citenamefont{Ruiz, Shapiro, and
  Tsokaros}}]{Ruiz2018}
\bibinfo{author}{\bibfnamefont{M.}~\bibnamefont{Ruiz}},
  \bibinfo{author}{\bibfnamefont{S.~L.} \bibnamefont{Shapiro}},
  \bibnamefont{and} \bibinfo{author}{\bibfnamefont{A.}~\bibnamefont{Tsokaros}},
  \bibinfo{journal}{Phys. Rev.} \textbf{\bibinfo{volume}{D97}},
  \bibinfo{pages}{021501} (\bibinfo{year}{2018}), \eprint{1711.00473}.

\bibitem[{\citenamefont{Shibata et~al.}(2017)\citenamefont{Shibata,
  Fujibayashi, Hotokezaka, Kiuchi, Kyutoku, Sekiguchi, and
  Tanaka}}]{Shibata2017}
\bibinfo{author}{\bibfnamefont{M.}~\bibnamefont{Shibata}},
  \bibinfo{author}{\bibfnamefont{S.}~\bibnamefont{Fujibayashi}},
  \bibinfo{author}{\bibfnamefont{K.}~\bibnamefont{Hotokezaka}},
  \bibinfo{author}{\bibfnamefont{K.}~\bibnamefont{Kiuchi}},
  \bibinfo{author}{\bibfnamefont{K.}~\bibnamefont{Kyutoku}},
  \bibinfo{author}{\bibfnamefont{Y.}~\bibnamefont{Sekiguchi}},
  \bibnamefont{and} \bibinfo{author}{\bibfnamefont{M.}~\bibnamefont{Tanaka}},
  \bibinfo{journal}{Phys. Rev.} \textbf{\bibinfo{volume}{D96}},
  \bibinfo{pages}{123012} (\bibinfo{year}{2017}), \eprint{1710.07579}.

\bibitem[{\citenamefont{Most et~al.}(2018)\citenamefont{Most, Weih, Rezzolla,
  and Schaffner-Bielich}}]{Most2018}
\bibinfo{author}{\bibfnamefont{E.~R.} \bibnamefont{Most}},
  \bibinfo{author}{\bibfnamefont{L.~R.} \bibnamefont{Weih}},
  \bibinfo{author}{\bibfnamefont{L.}~\bibnamefont{Rezzolla}}, \bibnamefont{and}
  \bibinfo{author}{\bibfnamefont{J.}~\bibnamefont{Schaffner-Bielich}}
  (\bibinfo{year}{2018}), \eprint{1803.00549}.

\bibitem[{\citenamefont{{Yagi} and {Yunes}}(2013{\natexlab{a}})}]{Yagi2013}
\bibinfo{author}{\bibfnamefont{K.}~\bibnamefont{{Yagi}}} \bibnamefont{and}
  \bibinfo{author}{\bibfnamefont{N.}~\bibnamefont{{Yunes}}},
  \bibinfo{journal}{Science} \textbf{\bibinfo{volume}{341}},
  \bibinfo{pages}{365} (\bibinfo{year}{2013}{\natexlab{a}}).

\bibitem[{\citenamefont{{Yagi} and {Yunes}}(2013{\natexlab{b}})}]{Yagi2013a}
\bibinfo{author}{\bibfnamefont{K.}~\bibnamefont{{Yagi}}} \bibnamefont{and}
  \bibinfo{author}{\bibfnamefont{N.}~\bibnamefont{{Yunes}}},
  \bibinfo{journal}{\prd} \textbf{\bibinfo{volume}{88}}, \bibinfo{eid}{023009}
  (\bibinfo{year}{2013}{\natexlab{b}}).

\bibitem[{\citenamefont{{Doneva} et~al.}(2015)\citenamefont{{Doneva},
  {Yazadjiev}, and {Kokkotas}}}]{Doneva2015}
\bibinfo{author}{\bibfnamefont{D.~D.} \bibnamefont{{Doneva}}},
  \bibinfo{author}{\bibfnamefont{S.~S.} \bibnamefont{{Yazadjiev}}},
  \bibnamefont{and} \bibinfo{author}{\bibfnamefont{K.~D.}
  \bibnamefont{{Kokkotas}}}, \bibinfo{journal}{\prd}
  \textbf{\bibinfo{volume}{92}}, \bibinfo{eid}{064015} (\bibinfo{year}{2015}),
  \eprint{1507.00378}.

\bibitem[{\citenamefont{{Doneva} et~al.}(2014)\citenamefont{{Doneva},
  {Yazadjiev}, {Staykov}, and {Kokkotas}}}]{Doneva2014a}
\bibinfo{author}{\bibfnamefont{D.~D.} \bibnamefont{{Doneva}}},
  \bibinfo{author}{\bibfnamefont{S.~S.} \bibnamefont{{Yazadjiev}}},
  \bibinfo{author}{\bibfnamefont{K.~V.} \bibnamefont{{Staykov}}},
  \bibnamefont{and} \bibinfo{author}{\bibfnamefont{K.~D.}
  \bibnamefont{{Kokkotas}}}, \bibinfo{journal}{\prd}
  \textbf{\bibinfo{volume}{90}}, \bibinfo{eid}{104021} (\bibinfo{year}{2014}),
  \eprint{1408.1641}.

\bibitem[{\citenamefont{{Kleihaus} et~al.}(2014)\citenamefont{{Kleihaus},
  {Kunz}, and {Mojica}}}]{Kleihaus2014}
\bibinfo{author}{\bibfnamefont{B.}~\bibnamefont{{Kleihaus}}},
  \bibinfo{author}{\bibfnamefont{J.}~\bibnamefont{{Kunz}}}, \bibnamefont{and}
  \bibinfo{author}{\bibfnamefont{S.}~\bibnamefont{{Mojica}}},
  \bibinfo{journal}{\prd} \textbf{\bibinfo{volume}{90}}, \bibinfo{eid}{061501}
  (\bibinfo{year}{2014}), \eprint{1407.6884}.

\bibitem[{\citenamefont{{Breu} and {Rezzolla}}(2016)}]{Breu2016}
\bibinfo{author}{\bibfnamefont{C.}~\bibnamefont{{Breu}}} \bibnamefont{and}
  \bibinfo{author}{\bibfnamefont{L.}~\bibnamefont{{Rezzolla}}},
  \bibinfo{journal}{\mnras} \textbf{\bibinfo{volume}{459}},
  \bibinfo{pages}{646} (\bibinfo{year}{2016}), \eprint{1601.06083}.

\bibitem[{\citenamefont{{Staykov} et~al.}(2016)\citenamefont{{Staykov},
  {Doneva}, and {Yazadjiev}}}]{Staykov2016}
\bibinfo{author}{\bibfnamefont{K.~V.} \bibnamefont{{Staykov}}},
  \bibinfo{author}{\bibfnamefont{D.~D.} \bibnamefont{{Doneva}}},
  \bibnamefont{and} \bibinfo{author}{\bibfnamefont{S.~S.}
  \bibnamefont{{Yazadjiev}}}, \bibinfo{journal}{\prd}
  \textbf{\bibinfo{volume}{93}}, \bibinfo{eid}{084010} (\bibinfo{year}{2016}),
  \eprint{1602.00504}.

\bibitem[{\citenamefont{Murray and Dermott}(1999)}]{Murray1999}
\bibinfo{author}{\bibfnamefont{C.~D.} \bibnamefont{Murray}} \bibnamefont{and}
  \bibinfo{author}{\bibfnamefont{S.~F.} \bibnamefont{Dermott}},
  \emph{\bibinfo{title}{Solar system dynamics}} (\bibinfo{publisher}{Cambridge
  university press}, \bibinfo{year}{1999}).

\bibitem[{\citenamefont{Poisson and Will}(2014)}]{Poisson2014}
\bibinfo{author}{\bibfnamefont{E.}~\bibnamefont{Poisson}} \bibnamefont{and}
  \bibinfo{author}{\bibfnamefont{C.~M.} \bibnamefont{Will}},
  \emph{\bibinfo{title}{Gravity: Newtonian, Post-Newtonian, Relativistic}}
  (\bibinfo{publisher}{Cambridge University Press}, \bibinfo{year}{2014}).

\bibitem[{\citenamefont{{Flanagan} and {Hinderer}}(2008)}]{Flanagan2008}
\bibinfo{author}{\bibfnamefont{{\'E}.~{\'E}.} \bibnamefont{{Flanagan}}}
  \bibnamefont{and}
  \bibinfo{author}{\bibfnamefont{T.}~\bibnamefont{{Hinderer}}},
  \bibinfo{journal}{\prd} \textbf{\bibinfo{volume}{77}}, \bibinfo{eid}{021502}
  (\bibinfo{year}{2008}).

\bibitem[{\citenamefont{Hinderer}(2008)}]{Hinderer2008}
\bibinfo{author}{\bibfnamefont{T.}~\bibnamefont{Hinderer}},
  \bibinfo{journal}{Astrophys. J.} \textbf{\bibinfo{volume}{677}},
  \bibinfo{pages}{1216} (\bibinfo{year}{2008}), \eprint{0711.2420}.

\bibitem[{\citenamefont{Binnington and Poisson}(2009)}]{Binnington2009}
\bibinfo{author}{\bibfnamefont{T.}~\bibnamefont{Binnington}} \bibnamefont{and}
  \bibinfo{author}{\bibfnamefont{E.}~\bibnamefont{Poisson}},
  \bibinfo{journal}{Phys. Rev.} \textbf{\bibinfo{volume}{D80}},
  \bibinfo{pages}{084018} (\bibinfo{year}{2009}), \eprint{0906.1366}.

\bibitem[{\citenamefont{Damour and Nagar}(2009)}]{Damour2009}
\bibinfo{author}{\bibfnamefont{T.}~\bibnamefont{Damour}} \bibnamefont{and}
  \bibinfo{author}{\bibfnamefont{A.}~\bibnamefont{Nagar}},
  \bibinfo{journal}{Phys. Rev.} \textbf{\bibinfo{volume}{D80}},
  \bibinfo{pages}{084035} (\bibinfo{year}{2009}), \eprint{0906.0096}.

\bibitem[{\citenamefont{{Hinderer} et~al.}(2010)\citenamefont{{Hinderer},
  {Lackey}, {Lang}, and {Read}}}]{Hinderer2010}
\bibinfo{author}{\bibfnamefont{T.}~\bibnamefont{{Hinderer}}},
  \bibinfo{author}{\bibfnamefont{B.~D.} \bibnamefont{{Lackey}}},
  \bibinfo{author}{\bibfnamefont{R.~N.} \bibnamefont{{Lang}}},
  \bibnamefont{and} \bibinfo{author}{\bibfnamefont{J.~S.}
  \bibnamefont{{Read}}}, \bibinfo{journal}{\prd} \textbf{\bibinfo{volume}{81}},
  \bibinfo{eid}{123016} (\bibinfo{year}{2010}).

\bibitem[{\citenamefont{Vines et~al.}(2011)\citenamefont{Vines, Flanagan, and
  Hinderer}}]{Vines2011}
\bibinfo{author}{\bibfnamefont{J.}~\bibnamefont{Vines}},
  \bibinfo{author}{\bibfnamefont{E.~E.} \bibnamefont{Flanagan}},
  \bibnamefont{and} \bibinfo{author}{\bibfnamefont{T.}~\bibnamefont{Hinderer}},
  \bibinfo{journal}{Phys. Rev.} \textbf{\bibinfo{volume}{D83}},
  \bibinfo{pages}{084051} (\bibinfo{year}{2011}), \eprint{1101.1673}.

\bibitem[{\citenamefont{Damour et~al.}(2012)\citenamefont{Damour, Nagar, and
  Villain}}]{Damour2012}
\bibinfo{author}{\bibfnamefont{T.}~\bibnamefont{Damour}},
  \bibinfo{author}{\bibfnamefont{A.}~\bibnamefont{Nagar}}, \bibnamefont{and}
  \bibinfo{author}{\bibfnamefont{L.}~\bibnamefont{Villain}},
  \bibinfo{journal}{Phys. Rev.} \textbf{\bibinfo{volume}{D85}},
  \bibinfo{pages}{123007} (\bibinfo{year}{2012}), \eprint{1203.4352}.

\bibitem[{\citenamefont{Del~Pozzo et~al.}(2013)\citenamefont{Del~Pozzo, Li,
  Agathos, Van Den~Broeck, and Vitale}}]{DelPozzo2013}
\bibinfo{author}{\bibfnamefont{W.}~\bibnamefont{Del~Pozzo}},
  \bibinfo{author}{\bibfnamefont{T.~G.~F.} \bibnamefont{Li}},
  \bibinfo{author}{\bibfnamefont{M.}~\bibnamefont{Agathos}},
  \bibinfo{author}{\bibfnamefont{C.}~\bibnamefont{Van Den~Broeck}},
  \bibnamefont{and} \bibinfo{author}{\bibfnamefont{S.}~\bibnamefont{Vitale}},
  \bibinfo{journal}{Phys. Rev. Lett.} \textbf{\bibinfo{volume}{111}},
  \bibinfo{pages}{071101} (\bibinfo{year}{2013}), \eprint{1307.8338}.

\bibitem[{\citenamefont{Maselli et~al.}(2013)\citenamefont{Maselli, Gualtieri,
  and Ferrari}}]{Maselli2013a}
\bibinfo{author}{\bibfnamefont{A.}~\bibnamefont{Maselli}},
  \bibinfo{author}{\bibfnamefont{L.}~\bibnamefont{Gualtieri}},
  \bibnamefont{and} \bibinfo{author}{\bibfnamefont{V.}~\bibnamefont{Ferrari}},
  \bibinfo{journal}{Phys. Rev.} \textbf{\bibinfo{volume}{D88}},
  \bibinfo{pages}{104040} (\bibinfo{year}{2013}), \eprint{1310.5381}.

\bibitem[{\citenamefont{Hinderer et~al.}(2016)}]{Hinderer2016}
\bibinfo{author}{\bibfnamefont{T.}~\bibnamefont{Hinderer}}
  \bibnamefont{et~al.}, \bibinfo{journal}{Phys. Rev. Lett.}
  \textbf{\bibinfo{volume}{116}}, \bibinfo{pages}{181101}
  (\bibinfo{year}{2016}), \eprint{1602.00599}.

\bibitem[{\citenamefont{Cardoso et~al.}(2017)\citenamefont{Cardoso, Franzin,
  Maselli, Pani, and Raposo}}]{Cardoso2017}
\bibinfo{author}{\bibfnamefont{V.}~\bibnamefont{Cardoso}},
  \bibinfo{author}{\bibfnamefont{E.}~\bibnamefont{Franzin}},
  \bibinfo{author}{\bibfnamefont{A.}~\bibnamefont{Maselli}},
  \bibinfo{author}{\bibfnamefont{P.}~\bibnamefont{Pani}}, \bibnamefont{and}
  \bibinfo{author}{\bibfnamefont{G.}~\bibnamefont{Raposo}},
  \bibinfo{journal}{Phys. Rev.} \textbf{\bibinfo{volume}{D95}},
  \bibinfo{pages}{084014} (\bibinfo{year}{2017}), \bibinfo{note}{[Addendum:
  Phys. Rev.D95,no.8,089901(2017)]}, \eprint{1701.01116}.

\bibitem[{\citenamefont{Sotiriou and Faraoni}(2010)}]{Sotiriou2010}
\bibinfo{author}{\bibfnamefont{T.~P.} \bibnamefont{Sotiriou}} \bibnamefont{and}
  \bibinfo{author}{\bibfnamefont{V.}~\bibnamefont{Faraoni}},
  \bibinfo{journal}{Rev.Mod.Phys.} \textbf{\bibinfo{volume}{82}},
  \bibinfo{pages}{451} (\bibinfo{year}{2010}).

\bibitem[{\citenamefont{De~Felice and Tsujikawa}(2010)}]{DeFelice2010a}
\bibinfo{author}{\bibfnamefont{A.}~\bibnamefont{De~Felice}} \bibnamefont{and}
  \bibinfo{author}{\bibfnamefont{S.}~\bibnamefont{Tsujikawa}},
  \bibinfo{journal}{Living Rev.Rel.} \textbf{\bibinfo{volume}{13}},
  \bibinfo{pages}{3} (\bibinfo{year}{2010}).

\bibitem[{\citenamefont{Nojiri and Odintsov}(2011)}]{Nojiri2011}
\bibinfo{author}{\bibfnamefont{S.}~\bibnamefont{Nojiri}} \bibnamefont{and}
  \bibinfo{author}{\bibfnamefont{S.~D.} \bibnamefont{Odintsov}},
  \bibinfo{journal}{Phys.Rept.} \textbf{\bibinfo{volume}{505}},
  \bibinfo{pages}{59} (\bibinfo{year}{2011}).

\bibitem[{\citenamefont{Birrell and Davies}(1984)}]{Birrell1984}
\bibinfo{author}{\bibfnamefont{N.~D.} \bibnamefont{Birrell}} \bibnamefont{and}
  \bibinfo{author}{\bibfnamefont{P.~C.~W.} \bibnamefont{Davies}},
  \emph{\bibinfo{title}{Quantum fields in curved space}}, \bibinfo{number}{7}
  (\bibinfo{publisher}{Cambridge university press}, \bibinfo{year}{1984}).

\bibitem[{\citenamefont{Babichev and Langlois}(2010)}]{Babichev2010}
\bibinfo{author}{\bibfnamefont{E.}~\bibnamefont{Babichev}} \bibnamefont{and}
  \bibinfo{author}{\bibfnamefont{D.}~\bibnamefont{Langlois}},
  \bibinfo{journal}{Phys.Rev.} \textbf{\bibinfo{volume}{D81}},
  \bibinfo{pages}{124051} (\bibinfo{year}{2010}).

\bibitem[{\citenamefont{Yazadjiev et~al.}(2014)\citenamefont{Yazadjiev, Doneva,
  Kokkotas, and Staykov}}]{Yazadjiev2014}
\bibinfo{author}{\bibfnamefont{S.~S.} \bibnamefont{Yazadjiev}},
  \bibinfo{author}{\bibfnamefont{D.~D.} \bibnamefont{Doneva}},
  \bibinfo{author}{\bibfnamefont{K.~D.} \bibnamefont{Kokkotas}},
  \bibnamefont{and} \bibinfo{author}{\bibfnamefont{K.~V.}
  \bibnamefont{Staykov}}, \bibinfo{journal}{JCAP}
  \textbf{\bibinfo{volume}{1406}}, \bibinfo{pages}{003} (\bibinfo{year}{2014}).

\bibitem[{\citenamefont{Astashenok et~al.}(2017)\citenamefont{Astashenok,
  Odintsov, and de~la Cruz-Dombriz}}]{Astashenok2017}
\bibinfo{author}{\bibfnamefont{A.~V.} \bibnamefont{Astashenok}},
  \bibinfo{author}{\bibfnamefont{S.~D.} \bibnamefont{Odintsov}},
  \bibnamefont{and} \bibinfo{author}{\bibfnamefont{A.}~\bibnamefont{de~la
  Cruz-Dombriz}}, \bibinfo{journal}{Class. Quant. Grav.}
  \textbf{\bibinfo{volume}{34}}, \bibinfo{pages}{205008}
  (\bibinfo{year}{2017}), \eprint{1704.08311}.

\bibitem[{\citenamefont{{Staykov} et~al.}(2014)\citenamefont{{Staykov},
  {Doneva}, {Yazadjiev}, and {Kokkotas}}}]{Staykov2014}
\bibinfo{author}{\bibfnamefont{K.~V.} \bibnamefont{{Staykov}}},
  \bibinfo{author}{\bibfnamefont{D.~D.} \bibnamefont{{Doneva}}},
  \bibinfo{author}{\bibfnamefont{S.~S.} \bibnamefont{{Yazadjiev}}},
  \bibnamefont{and} \bibinfo{author}{\bibfnamefont{K.~D.}
  \bibnamefont{{Kokkotas}}}, \bibinfo{journal}{\jcap}
  \textbf{\bibinfo{volume}{10}}, \bibinfo{eid}{006} (\bibinfo{year}{2014}),
  \eprint{1407.2180}.

\bibitem[{\citenamefont{{Yazadjiev} et~al.}(2015)\citenamefont{{Yazadjiev},
  {Doneva}, and {Kokkotas}}}]{Yazadjiev2015}
\bibinfo{author}{\bibfnamefont{S.~S.} \bibnamefont{{Yazadjiev}}},
  \bibinfo{author}{\bibfnamefont{D.~D.} \bibnamefont{{Doneva}}},
  \bibnamefont{and} \bibinfo{author}{\bibfnamefont{K.~D.}
  \bibnamefont{{Kokkotas}}}, \bibinfo{journal}{\prd}
  \textbf{\bibinfo{volume}{91}}, \bibinfo{eid}{084018} (\bibinfo{year}{2015}),
  \eprint{1501.04591}.

\bibitem[{\citenamefont{Sagunski et~al.}(2018)\citenamefont{Sagunski, Zhang,
  Johnson, Lehner, Sakellariadou, Liebling, Palenzuela, and
  Neilsen}}]{Sagunski2018}
\bibinfo{author}{\bibfnamefont{L.}~\bibnamefont{Sagunski}},
  \bibinfo{author}{\bibfnamefont{J.}~\bibnamefont{Zhang}},
  \bibinfo{author}{\bibfnamefont{M.~C.} \bibnamefont{Johnson}},
  \bibinfo{author}{\bibfnamefont{L.}~\bibnamefont{Lehner}},
  \bibinfo{author}{\bibfnamefont{M.}~\bibnamefont{Sakellariadou}},
  \bibinfo{author}{\bibfnamefont{S.~L.} \bibnamefont{Liebling}},
  \bibinfo{author}{\bibfnamefont{C.}~\bibnamefont{Palenzuela}},
  \bibnamefont{and} \bibinfo{author}{\bibfnamefont{D.}~\bibnamefont{Neilsen}},
  \bibinfo{journal}{Phys. Rev.} \textbf{\bibinfo{volume}{D97}},
  \bibinfo{pages}{064016} (\bibinfo{year}{2018}), \eprint{1709.06634}.

\bibitem[{\citenamefont{Naf and Jetzer}(2010)}]{Naf2010}
\bibinfo{author}{\bibfnamefont{J.}~\bibnamefont{Naf}} \bibnamefont{and}
  \bibinfo{author}{\bibfnamefont{P.}~\bibnamefont{Jetzer}},
  \bibinfo{journal}{Phys.Rev.} \textbf{\bibinfo{volume}{D81}},
  \bibinfo{pages}{104003} (\bibinfo{year}{2010}).

\bibitem[{\citenamefont{Yazadjiev and Doneva}(2015)}]{Yazadjiev:2015xsj}
\bibinfo{author}{\bibfnamefont{S.~S.} \bibnamefont{Yazadjiev}}
  \bibnamefont{and} \bibinfo{author}{\bibfnamefont{D.~D.} \bibnamefont{Doneva}}
  (\bibinfo{year}{2015}), \eprint{1512.05711}.

\bibitem[{\citenamefont{Antoniadis et~al.}(2013)\citenamefont{Antoniadis,
  Freire, Wex, Tauris, Lynch et~al.}}]{Antoniadis13}
\bibinfo{author}{\bibfnamefont{J.}~\bibnamefont{Antoniadis}},
  \bibinfo{author}{\bibfnamefont{P.~C.} \bibnamefont{Freire}},
  \bibinfo{author}{\bibfnamefont{N.}~\bibnamefont{Wex}},
  \bibinfo{author}{\bibfnamefont{T.~M.} \bibnamefont{Tauris}},
  \bibinfo{author}{\bibfnamefont{R.~S.} \bibnamefont{Lynch}},
  \bibnamefont{et~al.}, \bibinfo{journal}{Science}
  \textbf{\bibinfo{volume}{340}}, \bibinfo{pages}{6131} (\bibinfo{year}{2013}).

\bibitem[{\citenamefont{{Demorest} et~al.}(2010)\citenamefont{{Demorest},
  {Pennucci}, {Ransom}, {Roberts}, and {Hessels}}}]{Demorest10}
\bibinfo{author}{\bibfnamefont{P.~B.} \bibnamefont{{Demorest}}},
  \bibinfo{author}{\bibfnamefont{T.}~\bibnamefont{{Pennucci}}},
  \bibinfo{author}{\bibfnamefont{S.~M.} \bibnamefont{{Ransom}}},
  \bibinfo{author}{\bibfnamefont{M.~S.~E.} \bibnamefont{{Roberts}}},
  \bibnamefont{and} \bibinfo{author}{\bibfnamefont{J.~W.~T.}
  \bibnamefont{{Hessels}}}, \bibinfo{journal}{\nat}
  \textbf{\bibinfo{volume}{467}}, \bibinfo{pages}{1081} (\bibinfo{year}{2010}).

\bibitem[{\citenamefont{{Akmal} et~al.}(1998)\citenamefont{{Akmal},
  {Pandharipande}, and {Ravenhall}}}]{AkmalPR}
\bibinfo{author}{\bibfnamefont{A.}~\bibnamefont{{Akmal}}},
  \bibinfo{author}{\bibfnamefont{V.~R.} \bibnamefont{{Pandharipande}}},
  \bibnamefont{and} \bibinfo{author}{\bibfnamefont{D.~G.}
  \bibnamefont{{Ravenhall}}}, \bibinfo{journal}{\prc}
  \textbf{\bibinfo{volume}{58}}, \bibinfo{pages}{1804} (\bibinfo{year}{1998}).

\bibitem[{\citenamefont{{Douchin} and {Haensel}}(2001)}]{Douchin2001}
\bibinfo{author}{\bibfnamefont{F.}~\bibnamefont{{Douchin}}} \bibnamefont{and}
  \bibinfo{author}{\bibfnamefont{P.}~\bibnamefont{{Haensel}}},
  \bibinfo{journal}{\aap} \textbf{\bibinfo{volume}{380}}, \bibinfo{pages}{151}
  (\bibinfo{year}{2001}).

\bibitem[{\citenamefont{M?ther et~al.}(1987)\citenamefont{M?ther, Prakash, and
  Ainsworth}}]{Muether1987}
\bibinfo{author}{\bibfnamefont{H.}~\bibnamefont{M?ther}},
  \bibinfo{author}{\bibfnamefont{M.}~\bibnamefont{Prakash}}, \bibnamefont{and}
  \bibinfo{author}{\bibfnamefont{T.~L.} \bibnamefont{Ainsworth}},
  \bibinfo{journal}{Phys. Lett.} \textbf{\bibinfo{volume}{B199}},
  \bibinfo{pages}{469} (\bibinfo{year}{1987}).

\bibitem[{\citenamefont{Lattimer and Steiner}(2014)}]{Lattimer2014}
\bibinfo{author}{\bibfnamefont{J.~M.} \bibnamefont{Lattimer}} \bibnamefont{and}
  \bibinfo{author}{\bibfnamefont{A.~W.} \bibnamefont{Steiner}},
  \bibinfo{journal}{\apj} \textbf{\bibinfo{volume}{784}}, \bibinfo{eid}{123}
  (\bibinfo{year}{2014}), \eprint{1305.3242}.

\bibitem[{\citenamefont{{\"O}zel and Freire}(2016)}]{Oezel2016}
\bibinfo{author}{\bibfnamefont{F.}~\bibnamefont{{\"O}zel}} \bibnamefont{and}
  \bibinfo{author}{\bibfnamefont{P.}~\bibnamefont{Freire}},
  \bibinfo{journal}{Annu.~Rev.~Astron.~Astrophys.}
  \textbf{\bibinfo{volume}{54}}, \bibinfo{pages}{401} (\bibinfo{year}{2016}),
  \eprint{1603.02698}.

\end{thebibliography}

\end{document}